\documentclass[twocolumn,apj,appendixfloats,numberedappendix]{emulateapj-rtx4}

\usepackage{natbib}
\usepackage{graphicx}
\usepackage{amsmath,amssymb}
\usepackage{ulem}
\usepackage{enumitem}
\usepackage{txfonts}
\usepackage{epstopdf}

\newcommand{\chb}[1]{{#1}}

\newcommand{\be}{\begin{equation}}
\newcommand{\ee}{\end{equation}}
\newcommand{\gyrotropy}{$\mathrm{B_0}$-axisymmetry}
\newcommand{\gyrotropic}{$\mathrm{B_0}$-axisymmetric}
\newcommand{\gyro}{$\mathrm{B_0}$Axis}

\begin{document}

\title{Beyond the Maltese cross: geometry of turbulence between 0.2 and
1~AU}

\author{Andrea~Verdini}
\affil{UPMC, Paris, France \\ Lesia, Observatoire de Paris, Meudon, France \\ LPP, Ecole Polytechnique, Palaiseau, France}
\author{Roland~Grappin}
\affil{LPP, Ecole Polytechnique, Palaiseau, France.}


\date{\today}

\begin{abstract}
The spectral anisotropy of turbulent structures has been measured in the solar wind since 1990, relying on the assumption of axisymmetry about the mean magnetic field, $B_0$.  However, several works indicate that this hypothesis might be partially wrong, thus raising two questions: (i) is it correct to interpret measurements at 1 AU (the so-called Maltese cross) in term of a sum of slab and 2D turbulence? (ii) what information is really contained in the Maltese cross?

We solve direct numerical simulations of the MHD equations including the transverse stretching exerted by the solar wind flow and study the genuine 3D anisotropy of turbulence as well as that one resulting  from the assumption of axisymmetry about $B_0$.

\chb{We show that the evolution of the turbulent spectrum from 0.2 to 1 AU depends strongly on its initial anisotropy. An axisymmetric spectrum with respect to $B_0$ keeps its axisymmetry, i.e., resists stretching perpendicular to radial, while an isotropic spectrum becomes essentially axisymmetric with respect to the radial direction.}

We conclude that close to the Sun, slow-wind turbulence has a spectrum that is axisymmetric around $B_0$ and the measured 2D component at 1 AU describes the real shape of turbulent structures.  On the contrary, fast-wind turbulence has a more isotropic spectrum at the source and becomes radially symmetric at 1 AU. Such structure is hidden by the symmetrization applied to the data that instead returns a slab geometry.

\end{abstract}
\maketitle

\section{Introduction}

In a pioneering paper, \citet{1990JGR....9520673M} obtained for the first time
an average picture of the turbulent structures in the solar wind 
by computing the autocorrelation of the interplanetary magnetic field
fluctuations in different directions with respect to the mean field ($B_0$). 
and assuming axisymmetry about $B_0$.
Considering that single spacecraft measurements allow only to explore the radial
structure of the fluctuations, obtaining the Maltese cross was a big progress,
as it revealed the multidimensional structure of turbulence. The two-dimensional
(2D) autocorrelation was made up of two lobes, one elongated along increments
parallel to mean field, and the other elongated along increments perpendicular
to it (hence the term ``Maltese cross''). This particular shape was 
interpreted in terms of a mixture of 2D fluctuations with wavevectors and 
fluctuations perpendicular to the mean field (2D component), and of waves with 
wavevectors parallel to it (slab component), respectively.
 
The hypothesis of axisymmetry about $B_0$ underlying the Maltese cross picture is fully justified 
for homogeneous turbulence by theoretical, experimental and numerical results 
\citep{1981PhFl...24..825M,1983JPlPh..29..525S,1986PhFl...29.2433G} which all 
indicate that the nonlinear cascade leading to a turbulent spectrum proceeds mainly 
in directions perpendicular to the mean magnetic field. 
However, in the solar wind, the mean field direction is not the only symmetry axis 
for turbulent structures. Theoretical and numerical evidence 
\citep{1973Ap&SS..20..267V,1993PhRvL..70.2190G,Grappin:1996ey,Dong:2014fi} indicate that the flow 
direction, i.e. the radial axis, also plays a role in shaping the symmetry of
the turbulent spectrum in the Fourier space.
Also, for fluctuations with frequencies between 3 and 10 hours \cite{Saur:1999gy} 
found that the best theoretical model fitting solar wind data
is a mixture of 2D turbulence with wavevectors lying in a plane perpendicular to
the mean field and a spectrum of wavevectors aligned with the radial, not aligned 
with the mean field.

The argument that explains why the radial axis also plays a role is simple: as a
plasma volume is advected by the solar wind, the large scale flow cannot be
eliminated by a Galilean transformation, because it is \textit{radial, not uniform}. 
Indeed, after such a transformation, there remains an expanding flow
transverse to the radial that leads to a transverse stretching of the plasma
volume: this stretching has several consequences, an important one being that it slows 
down nonlinear coupling, at least in directions perpendicular to the radial.
In principle, at small enough scales the axisymmetry about $B_0$ should be valid, since
nonlinear couplings should overcome the transverse stretching: in fact their time scale 
becomes smaller while the expansion time scale is scale-independent.
However, we will see in this paper that the situation is less simple and that the radial 
symmetry can prevail even at small scales.

More specifically, this paper aims at understanding when the hypothesis of
axisymmetry about $B_0$ and the associated
Maltese cross picture are valid or not, and, at the same time, at guessing 
the true initial properties of turbulence close to the Sun that could lead to the structures 
observed at 1 AU.
We shall use for that the expanding box model (EBM \citep{1993PhRvL..70.2190G}), which consists in magnetohydrodynamic 
(MHD) equations modified to include the effect of the large scale radial flow of the wind. 
\chb{The EBM equations describe the evolution of a plasma parcel advected by a radial, uniform radial wind. Its conditions of validity are the following: (i) the angular width of the plasma volume must be small in order to allow neglecting curvature terms (but see however \cite{Grappin:1996ey}); (ii) the radial extent of the domain must be small enough  to allow assuming homogeneity within the domain; (iii) heliocentric distance must be larger than, say, 0.1 AU to be able to neglect systematic large-scale variations of the solar wind speed with heliocentric distance.}
The EBM equations have been used recently with success 
\citep{Verdini:2015bx} to reproduce and fully explain the local anisotropy of turbulent 
structures measured in the solar wind by \cite{2012ApJ...758..120C}.
The term \textit{local} means that the anisotropy is measured in a frame attached to 
the \textit{local} mean magnetic field that varies both with scale and location.
Note, however, that \textit{local} anisotropy is not easily related to the standard anisotropy 
studied in the present paper that is defined in a fixed frame, independent of scale 
(see \cite{2012ApJ...750..103M}).

In a previous work, \cite{1998JGR...10323705G} attempted to reproduce the
Maltese cross via direct numerical simulations of MHD equations, 
thus without taking into account the large scale radial flow of the wind.
Using the hypothesis of axisymmetry about $B_0$, they were able
to find separately the two lobes by varying the initial conditions of their
runs. They obtained the 2D component for initial conditions corresponding to 2D
turbulence or to pressure balance structures \citep{1995JGR...100.1763C}, and
the slab component for initial conditions corresponding to unidirectional 
Alfv\'en waves with wavevectors quasi-parallel to the magnetic field. 
However a mixture of these initial conditions led to isotropic autocorrelation, 
so they concluded that the two lobes of the Maltese cross result from
a mixture of different solar wind states.
This is indeed the case, as was shown by \cite{2005ApJ...635L.181D} and 
subsequent works \citep{2008JGRA..11301106H,2009JGRA..114.7213W,Weygand:2011p3064}, which 
successfully isolated the slab component and the 2D component by partitioning 
the fluctuations in fast and slow streams, respectively.

\begin{figure}[t]
\begin{center}
\includegraphics [width=\linewidth]{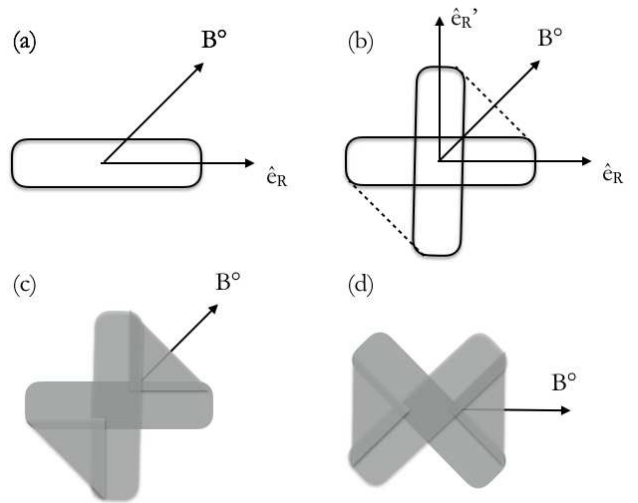}
\caption{Symmetrization around $B_0$ of a spectrum axisymmetric with respect to the radial direction, with the mean field $B_0$ at an angle $\pi/4$ with the radial.
(a) projection on the ($B_0,~e_R$) plane of an isocontour of the 3D spectrum that is axisymmetric around the radial direction;
(b) applying to the isocontour the assumption of axisymmetry about $B_0$;
(c) filling the interior of the contour to give an idea of the new
symmetrized spectrum;
(d) rotating the cartesian frame, with the $x$ axis aligned with the mean field $B_0$ as done when presenting the Maltese cross.
}
\label{fig1}
\end{center}
\end{figure}

Our purpose here is twofold: (i) to propose a description of the possible 
properties of turbulence close to the Sun compatible with these observations 
at 1 AU; (ii) explain how the hypothesis of axisymmetry about $B_0$ transforms this turbulence into the classical (2D, slab) model.
An extreme example of such a transformation is provided in fig.~\ref{fig1} 
which represents the effect of symmetrization around $B_0$ on a turbulent spectrum 
with radial symmetry, that is, an anisotropy completely ruled by expansion\footnote{We consider the 3D spectrum instead of the 3D autocorrelation
because rotating and averaging are more easily visualized in the Fourier space}.
The projection in the ecliptic plane of the 3D spectrum with radial
symmetry is shown in panel (a).
By averaging around the mean field direction, we obtain successively the panels (b) and (c).
This is equivalent to applying the hypothesis of axisymmetry about $B_0$ to measurements that
belong to several samples with different angles of the mean field with respect
to the radial.
Panel (d) represents the last step, i.e. the spectrum rotated in the frame associated with the mean field. 
The final spectrum has two properties: (i) by construction, it is axisymmetric with respect to the mean
field, while the true spectrum is axisymmetric with respect to the radial; (ii)
it has a complicated structure with main excitation along the mean field (i.e.
the observed slab component), which masks the true (physical) structure of the spectrum.

\begin{figure}
\begin{center}
\includegraphics [width=\linewidth]{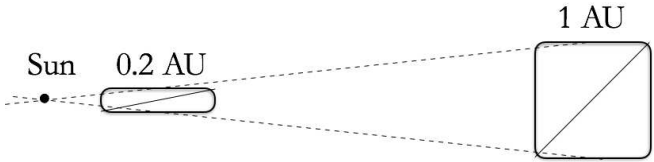}
\caption{Initial and final domains of simulation (and plasma volume as well) in the ecliptic plane. Thin lines: direction of mean magnetic field. For all runs with expansion, the aspect ratio of the domain varies from 1/5 to unity, and the mean magnetic field angle with the radial varies from $\tan^{-1}(1/5)$ to  $\tan^{-1}(1)=\pi/4$.
}
\label{f2}
\end{center}
\end{figure}

To reveal the possible initial structure and evolution of the (2D, radial slab)
two-component turbulence of \cite{Saur:1999gy}, we follow in this paper the
evolution of a plasma volume advected by the wind from $0.2$~AU to $1$~AU 
(fig.~\ref{f2}), using the EBM equations. 
The EBM equations have been used in \citet{Dong:2014fi} 
to explain basic properties of solar wind turbulence, namely
the anisotropy of the different components of fluctuations, both
kinetic and magnetic (also termed variance anisotropy). The present work extends this study by varying
(i) the initial conditions at 0.2 AU;
(ii) the ratio between the nonlinear turnover time based on the largest eddies and the linear stretching time.

The plan of the paper is as follows. 
Simulations and parameters are described in Section 2. Results on
the anisotropy of solar wind turbulence and its appearance in data under the
assumption of axisymmetry about the mean field are given in Section 3. 
In section 4 we present a discussion on the results and the impact of initial spectra and expansion
parameter on anisotropy. The last section contains the conclusions.

\section{Simulations and parameters}

The list of runs is indicated in Table~\ref{table1} along with the main
parameters. 
We now explain the different parameters.

\subsection{Expansion parameter $\epsilon$, time and distance}
In a plasma volume advected by the radial wind,
the linear stretching of the volume in directions perpendicular to the \textit{radial} 
transfers energy to \textit{larger scales} (fig.~\ref{f2}). 
On the other hand, the nonlinear couplings 
transfer energy to \textit{smaller scales} in directions perpendicular to the \textit{mean
field}. 
The relative strength of these different tendencies is quantified by the 
\textit{expansion parameter}, $\epsilon$, which is the ratio of the nonlinear time 
$t_{NL} = 1/(k_0 u)$ and the expansion time $t_e=R/U_0$:
\begin{eqnarray}
\epsilon = t_{NL}/t_e = (U_0/R) /(k_0 u)
\label{eps}
\end{eqnarray}
where $U_0$ is the (constant) wind speed, $R$ the heliocentric distance of the plasma volume, $k_0$ the minimum wavenumber associated with the dimension transverse (to radial) of the plasma volume, and $u$ the initial rms amplitude of the velocity fluctuations. 
Except if otherwise stated, the expansion parameter will be evaluated at the initial distance $R_0$: this is an important control parameter of each run.
Remark that the expansion parameter is not necessarily constant with time. In Table~\ref{table1},
we give both the initial value, $\epsilon$, and the value evaluated at the
end of the run, $\epsilon_{end}$: as one can see it increases for all runs.

A given run will be characterized by (i) the detailed initial conditions (see end of Section); (ii) the initial expansion parameter $\epsilon$.
Since the nonlinear couplings increase at small scales while the expansion
effect is scale-independent, one expects that for a given expansion rate $\epsilon$ of order unity
the wavenumber range is
made of two subsets: the larger scales are dominated by the linear effect of
expansion, while the smaller scales are dominated by nonlinear effects
\citep{Dong:2014fi}. 
However, we will find that the existence and location of such ranges 
depend largely on the anisotropy of the initial spectrum, and this will be a
basic result of the paper.

As time increases, the heliocentric distance increases as:
\begin{eqnarray}
R = R_0 + U_0 t
\end{eqnarray}
or, measuring time in terms of the initial nonlinear time:
\begin{eqnarray}
R/R_0 = 1 + \epsilon t
\end{eqnarray}
All expanding runs have $R_{max}=5R_0$, allowing us to follow the evolution
of the plasma between 0.2 and 1 AU, as stated in the introduction.

\subsection{Initial physical parameters}
The initial magnetic and kinetic fluctuations are solenoidal, obtained as a sum
of random-phase modes, and are at equipartition with root-mean-square value equal to one, 
$b_{rms}=u_{rms}=1$. Density and temperature are uniform; density is unity,
sound speed is $c_s\sim8$, so that the initial Mach number of the fluctuations
$M=u_{rms}/c_s=0.12$ and
remains small, as well as the relative amplitude of the compressible component.
\begin{table}
\caption[]{
List of runs and their parameters. All runs have initially $b_{rms}=u_{rms}=1$, density unity,
and end up with a domain of unit aspect ratio and a mean magnetic field $B_0$ at $45^0$ with the radial ($x$).
I.C. column: qualifies the initial spectrum, either ``\gyrotropic'' if
axisymmetric with respect to $B_0$, or ``ISO'' if isotropic.
$k_y^{cut}$ is the maximum vertical extent of the initial $k^{-1}$ spectrum (for runs with expansion).
In the column $B_0$ we indicate the $x$ and $y$ components of the initial mean magnetic field.
$\chi$ is a measure of the strength of turbulence based on initial conditions (see eq.~\ref{chi} and the corresponding text).
$\epsilon$ is the initial expansion parameter. 
$\epsilon_{end}$ is the final expansion parameter. 
$t_{end}$ is the end time, in nonlinear time units. 
All expanding runs have $R/R_0=5$ where $R_0,~R$ are the initial and final heliocentric distances, respectively.
$b_{rms}/B_0$ is the ratio between the rms and mean magnetic field at the end
of the simulation.}
\begin{tabular}{ccccccccc}
\multicolumn{5}{l}{   }      \\
Run &I.C. &$k_y^{cut}$&$B_0$&$\chi$&$\epsilon$ & $\epsilon_{end}$&$t_{end}$& $b_{rms}/B_0$\\
\hline
A &\gyro& &      $(\sqrt{2},\sqrt{2})/2$& 4&  0 & 0    & 4 & 0.8\\
B &\gyro& 128&       $(2,2/5)$& 2.5& 0.4 &     1.1  & 10& 0.6\\
C &ISO       & 128&  $(2,2/5)$& 1.2& 0.4 &     1.1  & 10& 0.6\\
\hline
D1 &\gyro   & 64&  $(2,2/5)$& 2.5& 0.4 &     0.97   & 10 & 0.5\\
D2 &\gyro   & 64&  $(2,2/5)$& 2.5& 1    &     1.9   & 4  & 0.5\\
D3 &\gyro   & 64&  $(2,2/5)$& 2.5& 2    &     3.3   & 2  & 0.5\\
D4 &\gyro   & 64&  $(2,2/5)$& 2.5& 3    &     4.8   & 4/3& 0.5\\
D5 &\gyro   & 64&  $(2,2/5)$& 2.5& 4    &     6.2   & 1  & 0.5\\
\hline
E1&ISO& 64&        $(2,2/5)$& 0.6& 0.1 &     0.26  & 40 & 0.6\\
E2&ISO& 64&        $(2,2/5)$& 0.6& 0.2 &     0.51  & 20 & 0.7\\
E3&ISO& 64&        $(2,2/5)$& 0.6& 0.4 &     1.0   & 10 & 0.5\\
E4&ISO& 64&        $(2,2/5)$& 0.6& 1   &     1.6   & 4  & 0.5\\
E5&ISO& 64&        $(2,2/5)$& 0.6& 2   &     2.8   & 2  & 0.5
\end{tabular}
\label{table1}
\end{table}

\subsection{Resolution, simulation domain and mean magnetic field}
The evolution of a turbulent spectrum in the solar wind is studied
by integrating the EBM equations with given initial conditions (decaying
simulations).
The resolution is $N_x=N_y=N_z=512$. 
Except for the single homogeneous run A with zero expansion ($\epsilon=0$), for all other runs the domain is expanding in the directions $y$ and $z$ perpendicular to the radial ($x$). 

The \textit{non-expanding} run A has an initial domain which is a cube with
sizes $L_x=L_y=L_z=2\pi$, and mean magnetic field $B_0=(1/\sqrt 2,1/\sqrt 2)$ in the $xOy$ plane.

The \textit{expanding} runs have an initial domain elongated by a factor 5 in the radial direction: 
$L_x= 5L_y=5L_z = 5 \times 2 \pi$. 
In doing so, we anticipate the stretching perpendicular to the radial, which transforms the domain into a cube at the distance $R=5R_0$.
Remark that one could start with a cubic domain as well, and thus end up with a
domain stretched in the directions perpendicular to radial (this was the choice
adopted in \citealt{Dong:2014fi}).
We prefer to end up with a cubic domain because (i) we are mostly interested in the turbulent state close to 1 AU; (ii) we think that nonlinear interactions are essentially local.
In this way we hope to better catch the properties of the turbulent cascade at the end of the simulation.

The initial mean magnetic field makes a small angle with the radial: $B_0=(2, 2/5)$. 
As the distance increases by a factor 5, the mean magnetic field rotates due to
magnetic flux conservation and makes an angle of $45^0$ with the radial at the
end of the simulation (see fig.~\ref{f2}).

\subsection{Initial spectrum}
For the homogeneous run A, we consider a bi-gaussian spectrum of the form
$\exp[-(k_\bot/2\Delta k_\bot)^2-(k_\|/2\Delta k_\|)^2]$ with a larger width
in directions perpendicular to the mean field, $\Delta k_\bot=4\Delta k_\|= 4$.
Note that changing the precise ratio, e.g., taking an isotropic initial spectrum, does not change the results at times longer than a couple of nonlinear times.

We now describe expanding runs.
We consider as in \cite{Dong:2014fi} an initial fluctuation spectrum at equipartition between magnetic and kinetic fluctuations, with a 1D $k^{-1}$ scaling, thus mimicking the fossil part of the spectrum measured in the fast streams. 
It is convenient as well to use such a strong small-scale excitation because otherwise,
when only large scales are present initially, a too large
expansion parameter prevents the direct cascade to form \citep{Dong:2014fi}.

\begin{figure}  
\begin{center}
\includegraphics [width=0.48\linewidth]{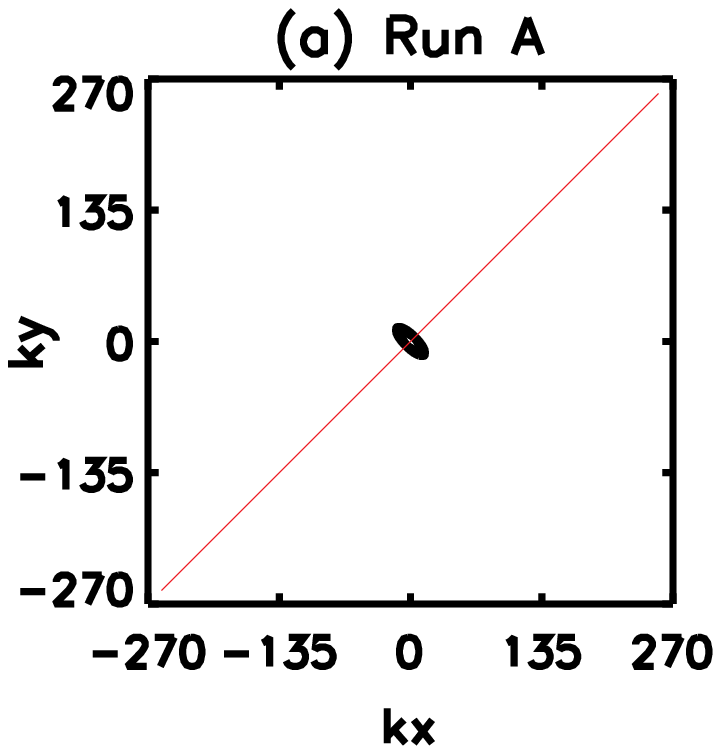}
\includegraphics [width=0.48\linewidth]{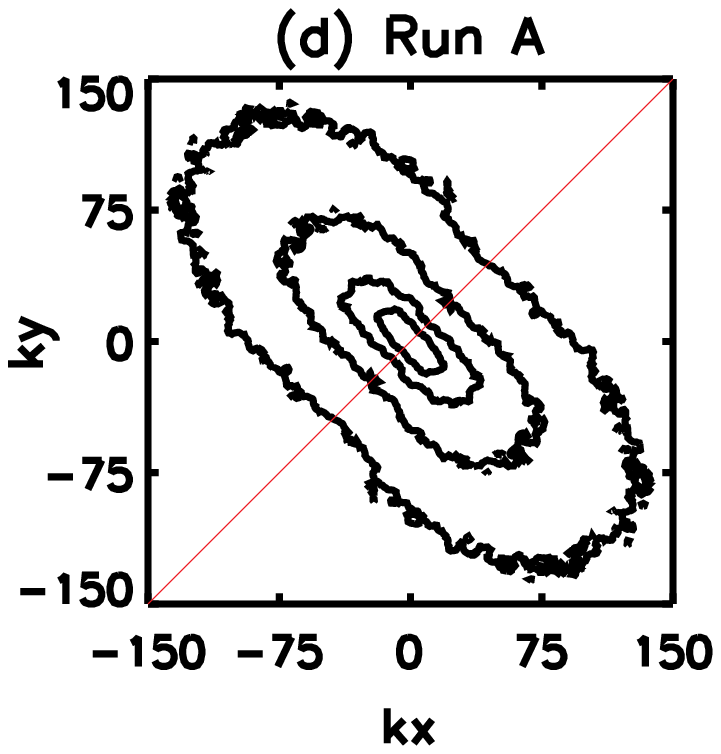}
\includegraphics [width=0.48\linewidth]{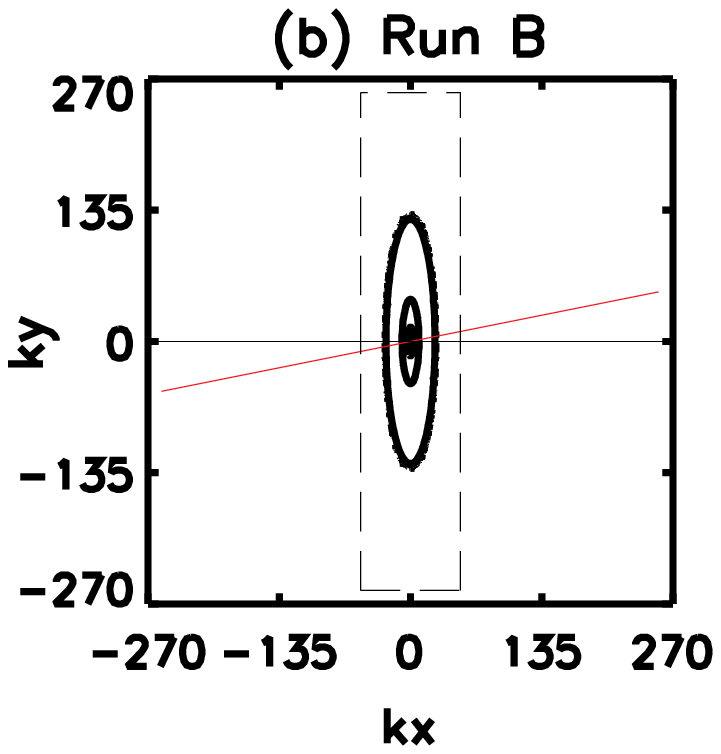}
\includegraphics [width=0.48\linewidth]{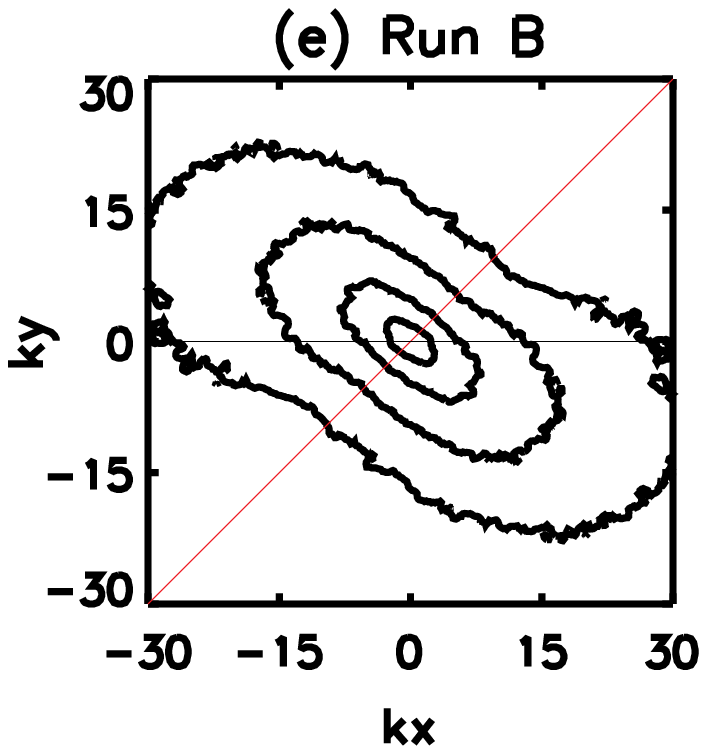}
\includegraphics [width=0.48\linewidth]{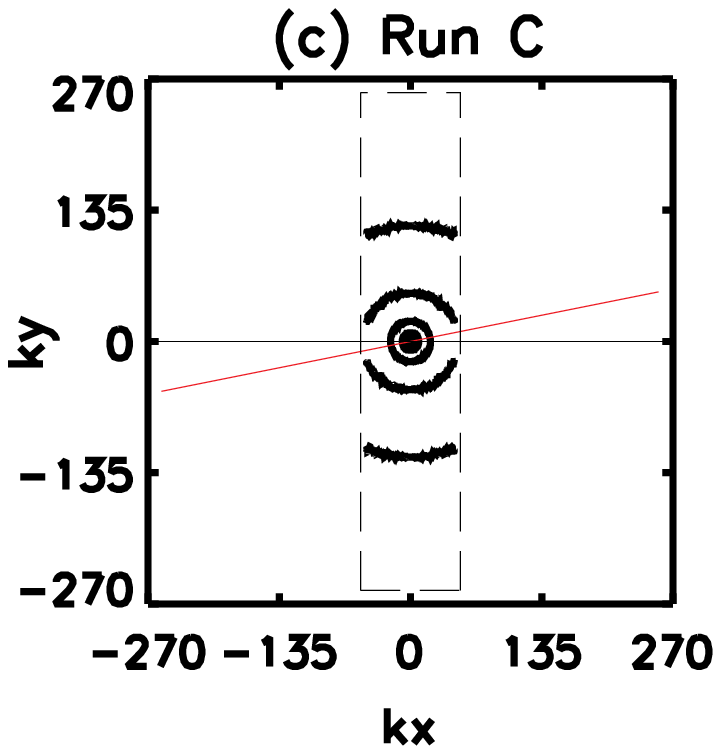}
\includegraphics [width=0.48\linewidth]{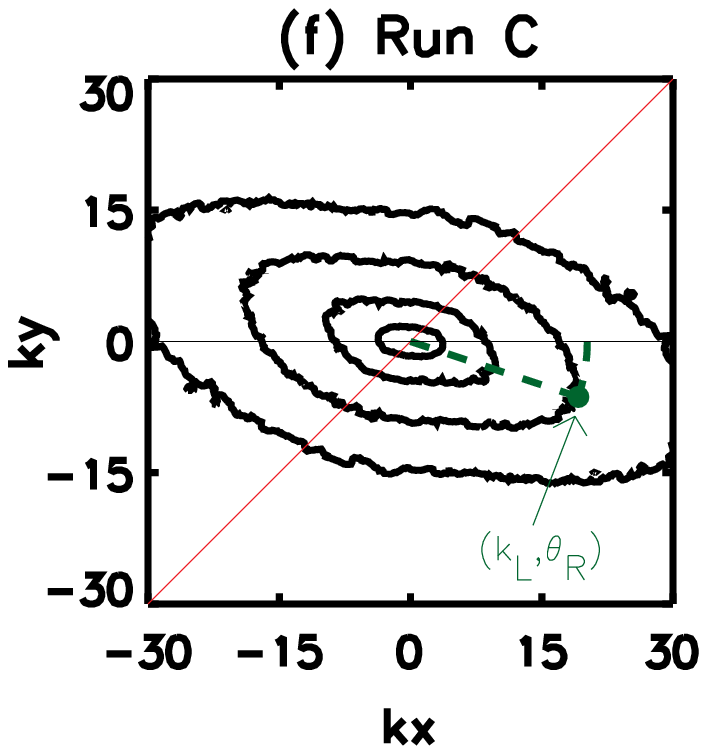}
\caption{Isocontours of the 3D magnetic energy spectrum in the ($k_x,k_y)$
plane, i.e., for $k_z=0$ for Runs A, B and C. Left panels, at $t=0$;
right panels at the end of the simulation (see
Table~\ref{table1}).
The oblique red line indicates the mean magnetic field direction.  
In panels (b) and (c) the Fourier domain at the initial distance $R=R_0 = 0.2$~AU
is bounded by the dashed horizontal and vertical lines.
In panels (e) and (f) only a part of the Fourier domain is shown at the final
distance of $R=1$~AU (the true extent is $[-51.2,51.2]^2$ with minimum wavenumber 0.2).
In panel (f) we illustrate the measure of the anisotropy that is used to
construct fig.~\ref{f7}: the filled dot indicated by the arrow marks the maximal distance of a given
isocontour level from the center in the polar coordinates
($k_L,\theta_R$).
}
\label{f3}
\end{center}
\end{figure}

We consider two variants of the $k^{-1}$ spectrum in figs.~\ref{f3}b,c.
First, we consider (panel b) energy isocontours elongated in the radial
direction, with an aspect ratio of 5, thus following the shape of the initial
domain in Fourier space (dashed lines), which is opposite to that in real space (fig.~\ref{f2}).
As in expanding runs the initial mean field forms a small angle with the radial
($tan^{-1}(1/5)$), this spectrum is approximately axisymmetric with respect to
$B_0$ with an aspect ratio roughly corresponding to a critical balance condition
(we checked that choosing true or approximate axisymmetry does not affect the results
presented here). 
Runs starting with such spectra are denoted by ``\gyro'' in
Table~\ref{table1}, and in the text ``\gyrotropic''. 
This term will denote at the same time axisymmetry with respect to the mean field and a spectrum principal axis perpendicular to $B_0$.
Of course, this property is not necessarily conserved with time.

Second, we considered energy isocontours with aspect ratio unity, thus not
following the shape of the initial plasma volume (see fig.~\ref{f3}c). Such runs
are denoted by ``ISO'' in Table~\ref{table1}, and in the text ``isotropic''. Again, this property is not necessarily conserved with time.

The $k^{-1}$ spectrum is cut to zero for $k_{y,z} \ge k_y^{cut}$ (the value is
128 or 64 depending on the run, see Table~\ref{table1}).
The aspect ratio of the truncation in wavevector space follows the aspect ratio of the isocontours
(isotropic or \gyrotropic).
Note that in practice, the \textit{isotropic} spectrum of run C is
truncated in the radial direction (x) by the domain boundary, not by the
truncation wavenumber $k_y^{cut}$ (see figs.~\ref{f3}b-c, in which the domain
boundaries are indicated by dashed horizontal and vertical lines). On the
contrary, for the class of isotropic runs $\mathrm{E_{1...5}}$ the truncation
is at a smaller wavenumber and the spectrum is almost truly isotropic (their
initial conditions are represented by the three inner circles in
fig.~\ref{f3}c).
In table~\ref{table1} we also indicate the parameter 
\be
\chi=B_0 k_\|^{cut}/b_{rms}k_\bot^{cut},
\label{chi}
\ee
with $k_\bot^{cut}$ and $k_\|^{cut}$ indicating the maximal wavenumber excited in the direction perpendicular and parallel to the mean field respectively (for expanding runs 
$k_\bot^{cut}\sim k_y^{cut}$ and $k_\|^{cut}\sim k_x^{cut}$). This parameter
complements the information on the symmetry of the spectrum since it quantifies
the strength of turbulence as the ratio of the smallest Alfv\'en time to the
smallest nonlinear time associated with initial conditions\footnote{The same
parameter was used in forced simulations to induce weak or strong turbulence regimes \citep{Dmitruk:1999p2192,Rappazzo:2007p454,Perez:2008p674,2012PhRvL.109b5004V}}. 
As a rule, all \gyrotropic{} runs have initially strong turbulence ($\chi>1$), while isotropic runs have weak turbulence ($\chi<1$). The only exception is run C that has $\chi\sim1$ due to the truncation by the domain boundary.

\section{Results}

\begin{figure*}[t]
\begin{center}
\includegraphics [width=0.32\linewidth]{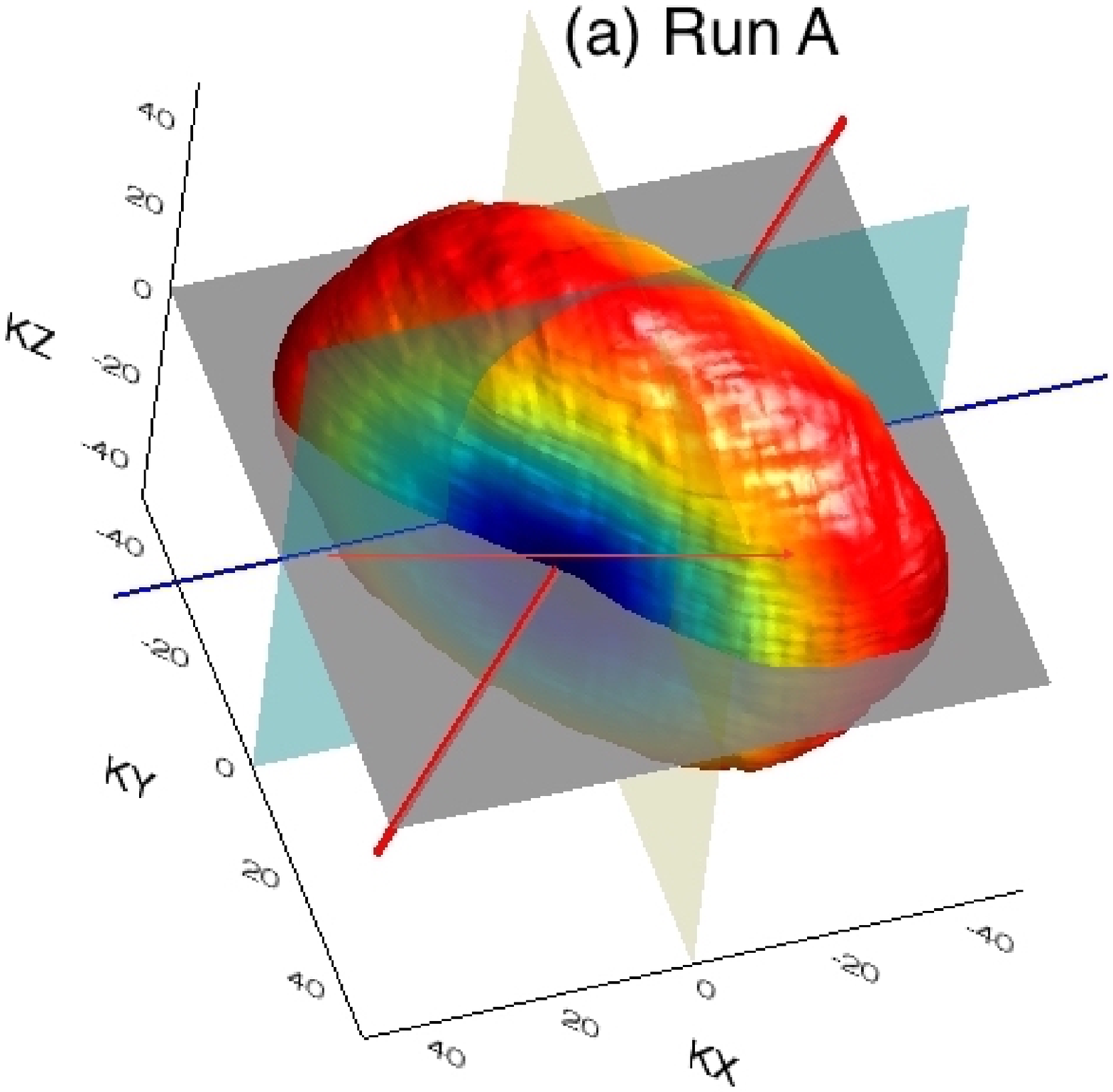}
\includegraphics [width=0.32\linewidth]{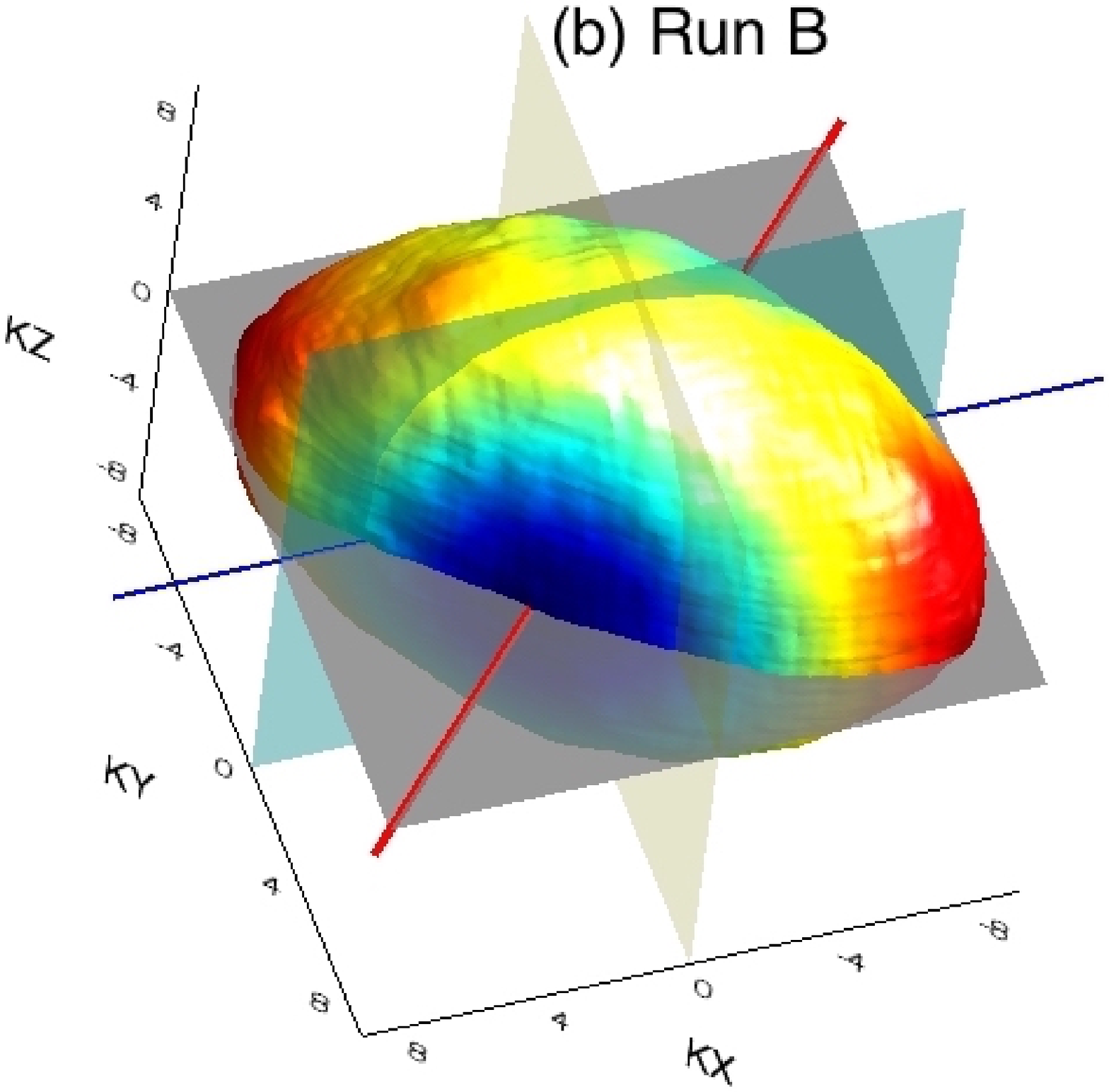}
\includegraphics [width=0.32\linewidth]{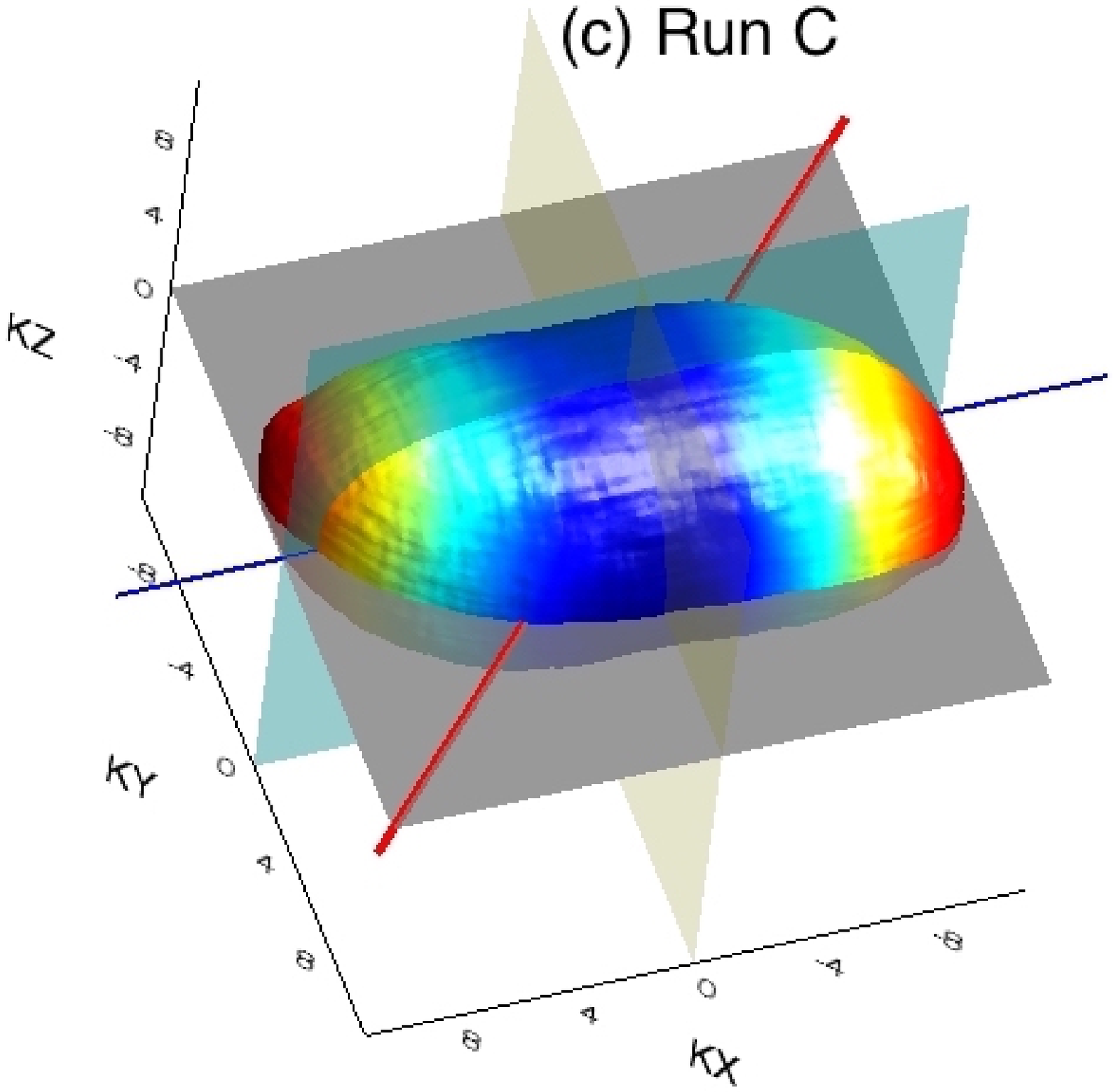}
\caption{Runs A, B, C.
Representative isosurface \chb{in the inertial range} of the 3D magnetic energy spectrum at 1 AU (end of each run, cf. Table~\ref{table1}).
Color indicates the distance to origin and is meant to give redundant information on the shape of the isosurface. 
}
\label{f4}
\end{center}
\end{figure*}

\begin{figure*}
\begin{center}
\includegraphics [width=0.32\linewidth]{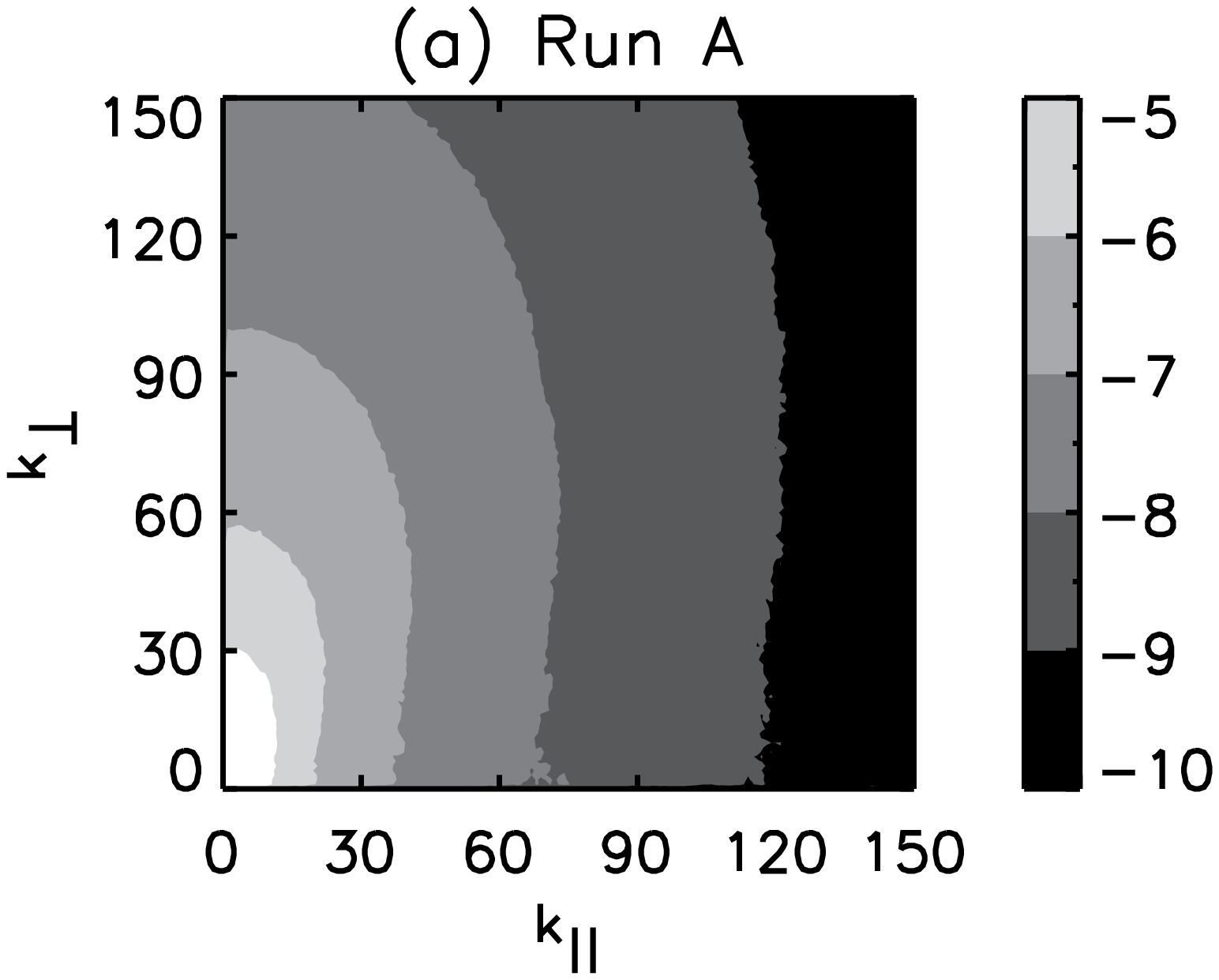}
\includegraphics [width=0.32\linewidth]{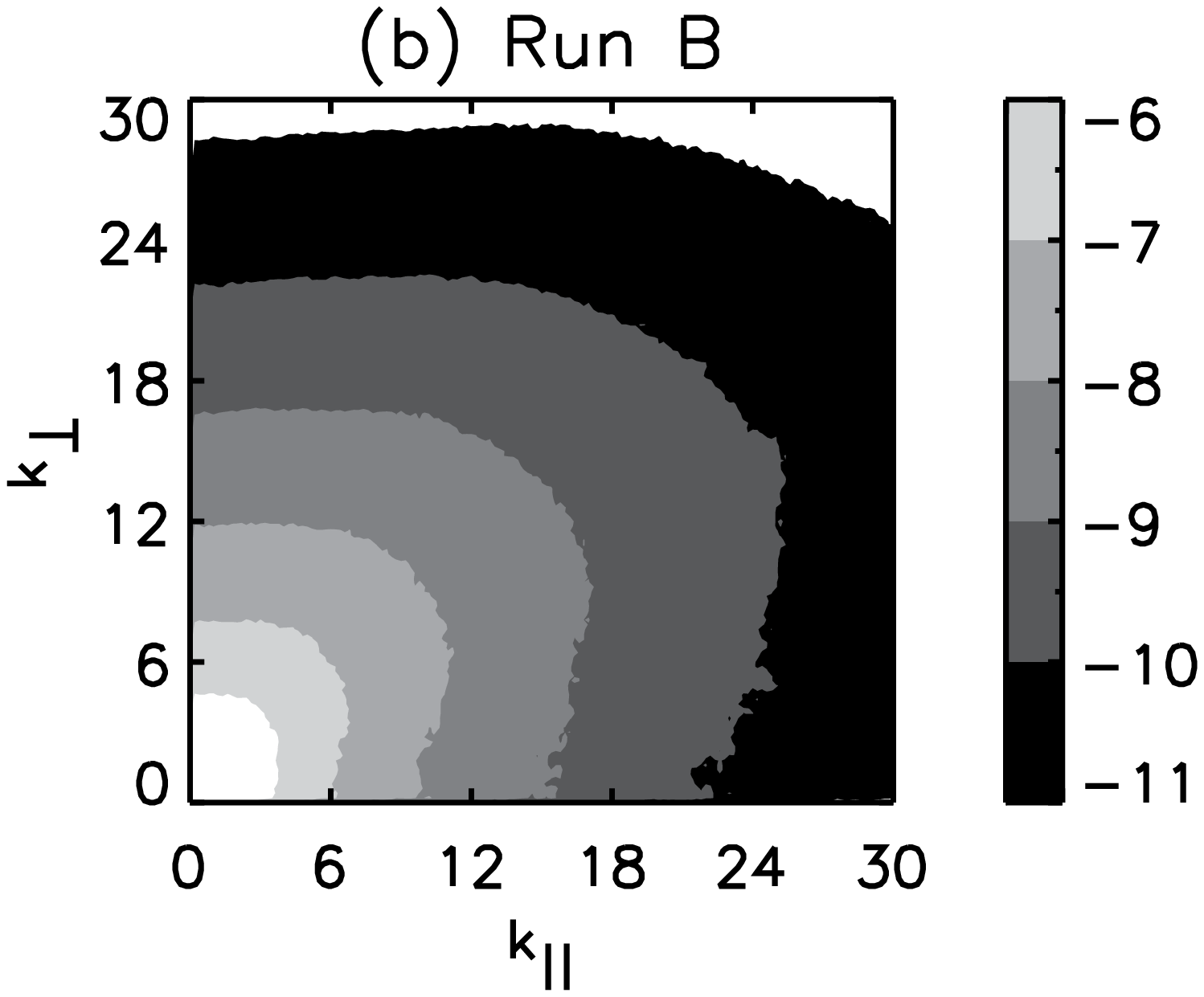}
\includegraphics [width=0.32\linewidth]{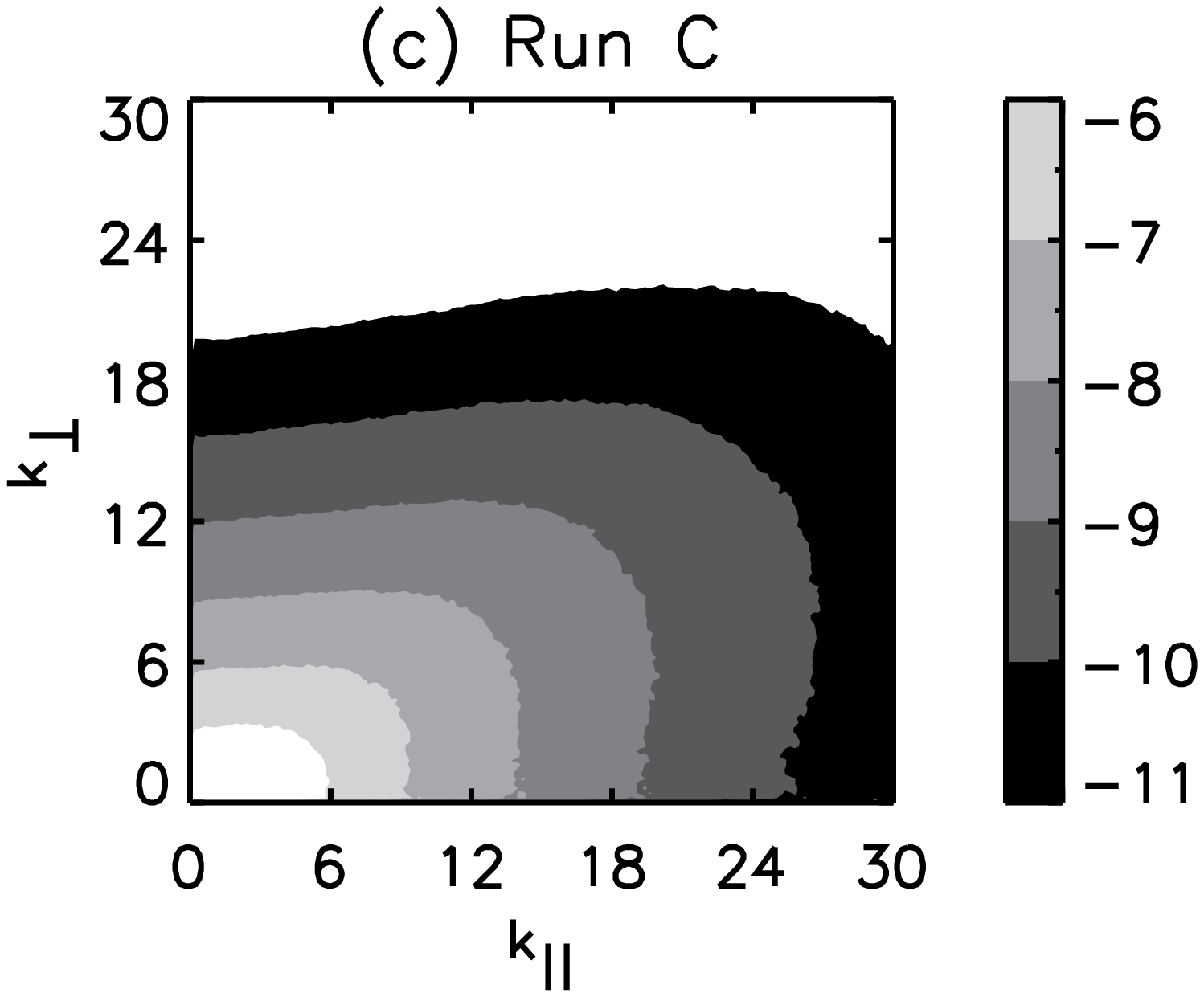}
\caption{Run A, B, C at the end of the simulation (see Table~\ref{table1}). 3D spectra, after
symmetrization around the mean magnetic field, in the frame attached to the mean field ($k_\|$, $k_\bot$) indicate the axis parallel 
and perpendicular to the mean field, respectively.
}
\label{f5}
\end{center}
\end{figure*}

We focus in this section on the first three runs of Table~\ref{table1}: run A without
expansion and \gyrotropic{} initial conditions, run B with expansion and
\gyrotropic{} initial conditions, run C with expansion and isotropic initial conditions.
Our analysis of the magnetic structure of each run will follow the following
steps.

First we represent 3D magnetic spectra at 0.2 and 1 AU 
with some detail in order to reveal the true evolution of anisotropy.

Second we draw ``$\mathrm{B_0}$-symmetrized" 3D spectra, obtained by averaging the 3D spectrum
over the azimuthal angle around the mean magnetic field axis,
\be
E_{3D}(k_\|,k_\bot)=1/2\pi\int E_{3D}(k_\|,k_\bot,\phi)\mathrm{d}\phi,
\ee 
in order to understand how much this procedure reveals or hides details about
the true 3D spectra.

Third, we transform the \gyrotropic{} 3D spectrum into the 2D autocorrelation,
\be
A(\ell_{||},\ell_\bot)=\int E_{3D}k_\bot
  \exp[i (k_\|\ell_\|+k_\bot\ell_\bot)]\mathrm{d}k_\|\mathrm{d}k_\bot,
\ee
to make a link between our simulations and the observed solar wind structures
(the Maltese cross). 


In this Section, the role of initial conditions is analyzed in runs B and C, at a fixed expansion parameter, with Run A (zero expansion) playing the role of a test simulation.
The role of the expansion parameter $\epsilon$ is analyzed later in the Discussion with runs 
$\mathrm{D_{1...5}}$ and $\mathrm{E_{1...5}}$.

\subsection{The 3D structure}

In the right panels of Figure~\ref{f3} we show isocontours of the \textit{ecliptic} cut (i.e., $k_z=0$) of the 3D spectra after four nonlinear times for run A, and ten nonlinear times for runs B and C (thus at $R=1$ AU).
The mean magnetic field is indicated by a red line, which has an angle $\pi/4$
with the radial direction ($k_x$, black line).
One sees that for run A the cascade proceeds perpendicularly to the mean field.
For run B, this is about true, with a small deviation in the radial direction.
For run C, the isocontours show an equal amount of stretching towards the radial
direction and the field-perpendicular direction.
\chb{Note that runs B and C develop a 1D spectrum with scaling close to $k^{-5/3}$ in the subrange $1 \le k \le 10$ (not shown)}

\begin{figure*}[t]
\begin{center}
\includegraphics [width=0.31\linewidth]{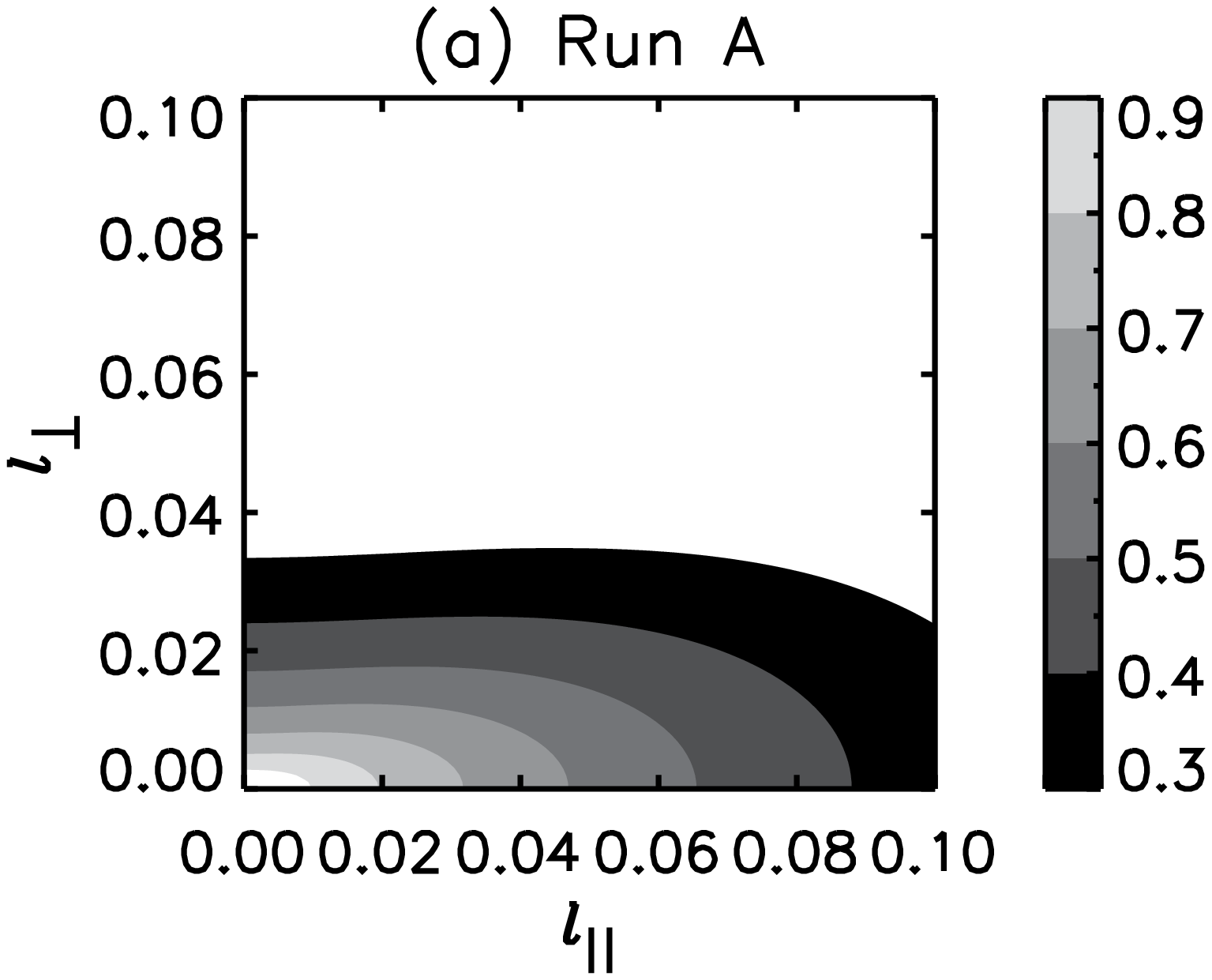}
\includegraphics [width=0.31\linewidth]{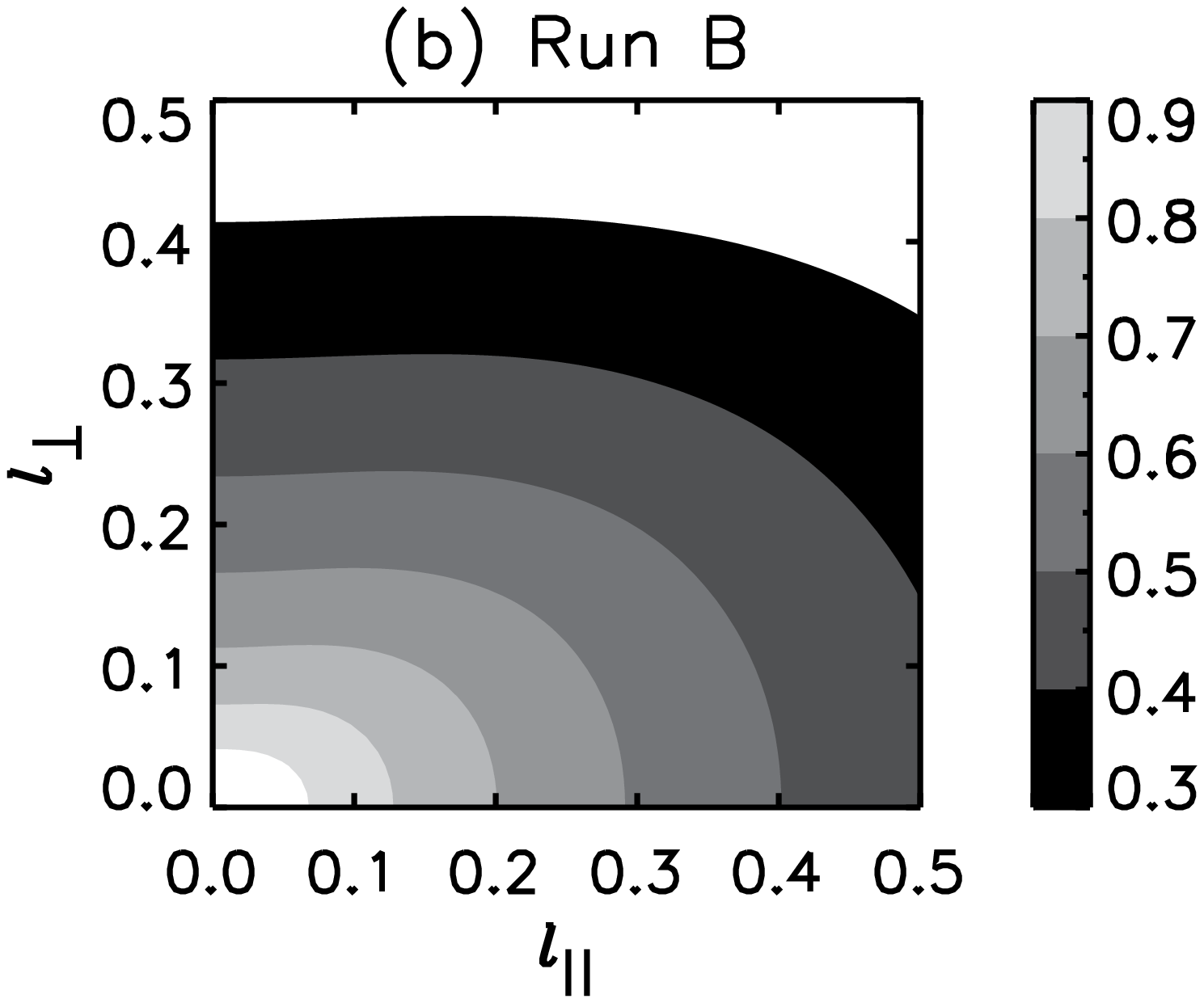}
\includegraphics [width=0.31\linewidth]{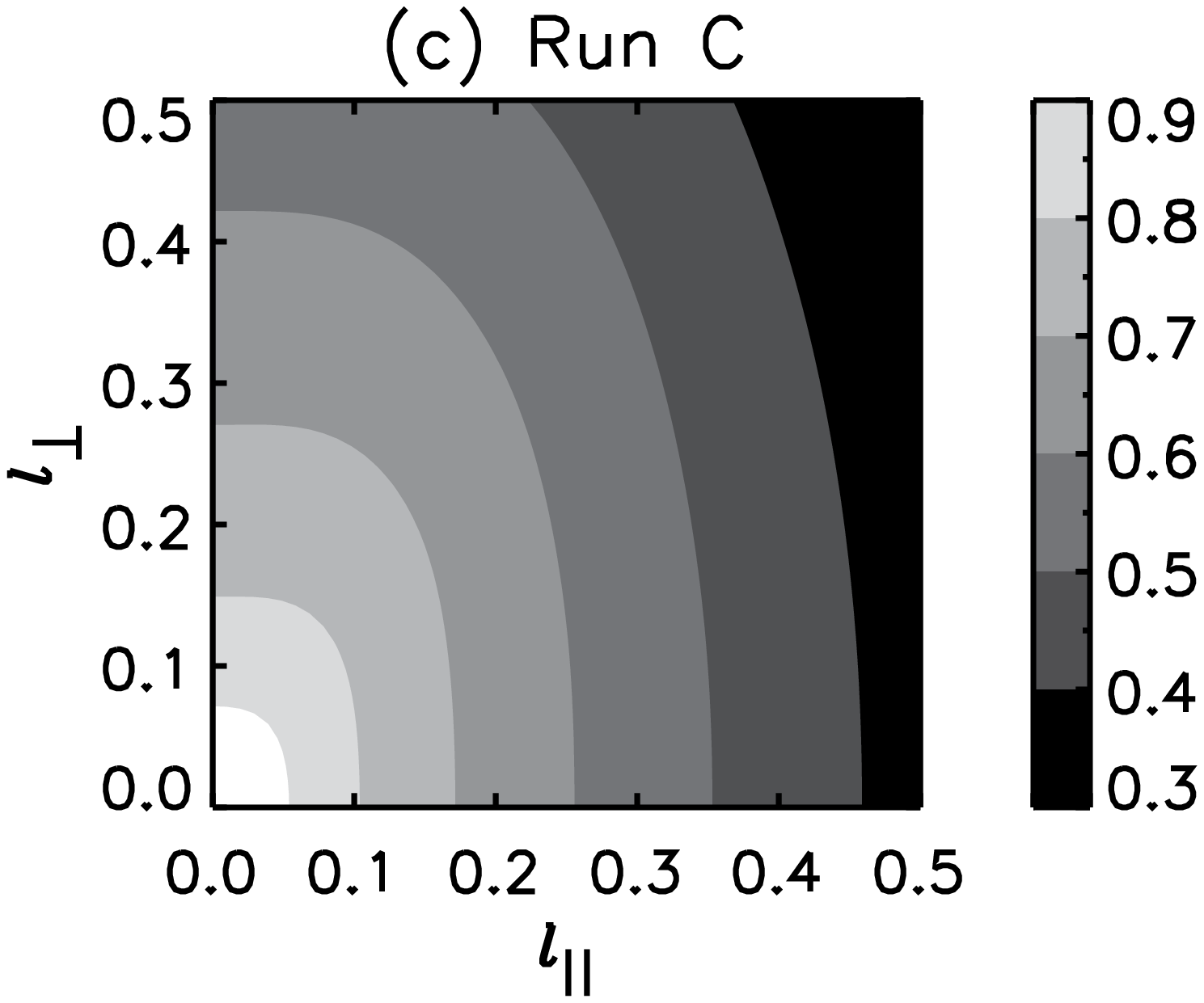}
\caption{Run A, B, C at the end of the simulation (see Table~\ref{table1}).
2D autocorrelation obtained from the \gyrotropic{} 3D spectra of fig.~\ref{f5} ($\ell_\|,~\ell_\bot$ indicate increments parallel and perpendicular to the mean field, respectively).
}
\label{f6}
\end{center}
\end{figure*}

The ecliptic view is complemented by fig.~\ref{f4} where we show a 3D perspective of one
representative isosurface of the spectrum taken in the inertial range, for the
three runs A, B, and C. Colors give the distance to the origin as a redundant information. 
The red diagonal line is the $B_0$ direction, the blue line indicates the radial $k_x$ direction.
Again, the two runs A and B appear to both exhibit axisymmetry with respect to
$B_0$ (with a cascade perpendicular to it), while run C shows a dominant radial
axisymmetry, 
with what resembles a cascade along the radial.
Note that for run B not only is the
symmetry axis slightly tilted with respect to the mean field but also 
axisymmetry is only roughly established (the isosurface is less elongated in the
$k_z$ direction than in the perpendicular direction lying in the ecliptic plane, $k_z=0$).

\subsection{Symmetrization around the mean field}
We now use ``blindly" the hypothesis of axisymmetry about $B_0$, that is, we average all 3D spectra on the azimuthal angle around the mean field.
The result is shown in fig.~\ref{f5}.
The dominant symmetry of the spectrum is respected for run A, in a mild way for run B, not at all for run C.
Indeed, the cascade is perpendicular to $B_0$ for run A.
For run B, the deviation from \gyrotropy{} is large enough, so that the
symmetrization transforms the spectrum into a quasi-isotropic spectrum.
For run C, the deviation from \gyrotropy{} is so large that the
symmetrization leads to a
spectrum elongated along the direction \textit{parallel to the mean field}.

The last step consists in transforming the averaged 3D spectra into 2D spectra
by integrating, and then taking the Fourier transform to recover the 2D
correlation figures, allowing comparison with 2D-autocorrelation of solar wind data.
The result is shown in fig.~\ref{f6}. 
Run A has a 2D-correlation elongated in the parallel direction. Run B has a
smaller elongation but still in the parallel direction while run C has its correlation elongated in the perpendicular direction.
These last two opposite elongations are strongly reminiscent of the two figures
obtained by the analysis of respectively slow and fast winds by
\cite{2005ApJ...635L.181D} that correspond to the two lobes of the
Maltese cross (in which fast and slow wind are mixed).

\section{Discussion}
\subsection{The 3D anisotropy and the Maltese cross}
We have examined three runs, run A with no expansion, two runs
with the same expansion rate but different initial conditions: a
\gyrotropic{} and anisotropic spectrum (run B) and an isotropic spectrum (run C).
We have two kinds of conclusions dealing respectively with the true structure of
turbulence and with its \textit{apparent} structure, i.e., after
symmetrization around $B_0$.

About the true structure of turbulence. Without expansion the end-result is \textit{always} that the spectrum is elongated in directions perpendicular to the mean field
\citep{1981PhFl...24..825M,1983JPlPh..29..525S,1986PhFl...29.2433G}.
However, with expansion, the conclusion is different: the ``end'' result depends strongly on the initial condition, as we have seen: the ``perpendicular'' cascade is not an attractor, or at least it is a weak attractor.

About the symmetrized structure, namely the 2D-correlation. In absence of
expansion it captures the true 3D anisotropy, as expected. In presence of
expansion, initial conditions ``perpendicular to mean field'' transform into
itself, that is the so-called 2D turbulence. In fact the true 3D anisotropy is
roughly axisymmetric with a symmetry axis close to the
mean field direction, so that the 2D correlation returns
qualitatively the correct anisotropy.
With expansion still, ``isotropic'' initial conditions transform into
the typical figure of the so-called ``slab" turbulence, with isocontour
elongated in the direction parallel to the mean field. In fact, because the true symmetry
axis is along the radial direction, the symmetrization changes qualitatively
the autocorrelation, as represented in Figure~\ref{fig1}, ultimately 
transforming it from slab along the radial direction into slab along the
mean-field direction.

In view of the observed association between (i) fast winds and slab signature;
(ii) slow winds and 2D signature, it is thus tempting to propose that turbulence
at the source of fast winds has an isotropic spectrum, and turbulence at the
source of slow winds has an anisotropic spectrum with a cascade
perpendicular to the mean field. We will come back to this point in the
conclusions.

\begin{figure*}[t]
\begin{center}
\includegraphics [width=0.48\linewidth]{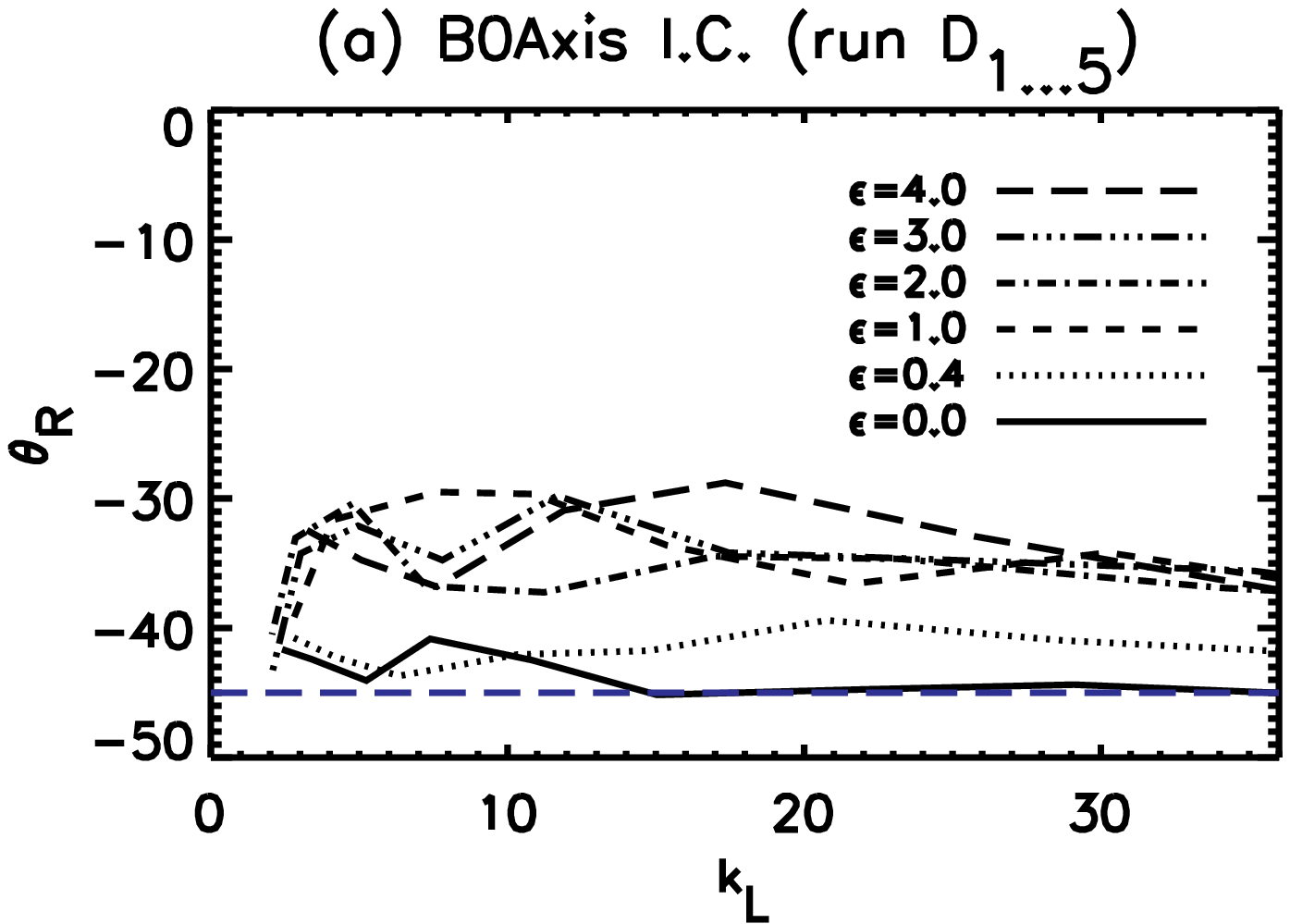}
\includegraphics [width=0.48\linewidth]{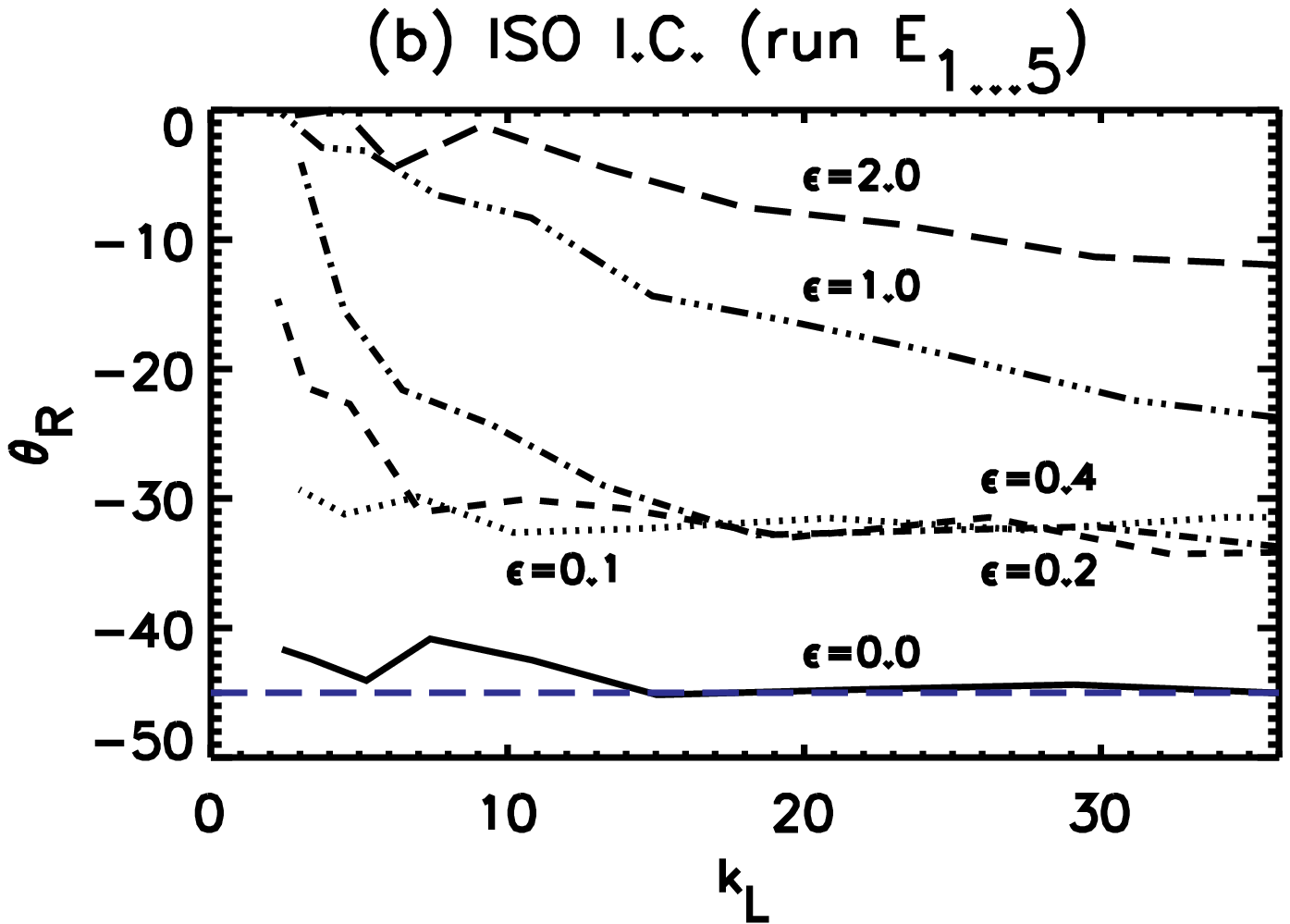}
\caption{Angle $\theta_R$ between the largest axis of isocontours and the radial
direction versus the lengths of the largest axis $k_L$ for different levels of isocontours (see fig.~\ref{f3}f for an example). 
(a) Runs $\mathrm{D_{1...5}}$ (\gyrotropic{} initial conditions); (b) Runs
$\mathrm{E_{1...5}}$ (isotropic initial conditions). The long-dashed line is the expected $B_0$-anisotropy at $-45^o$. 
\chb{The case of run A with $\epsilon=0$ (solid line) is added on both panels, with abscissa renormalized by a factor 5, thus taking into account the different domain sizes (see fig.~\ref{f3})}.
}
\label{f7}
\end{center}
\end{figure*}

\subsection{Anisotropy scaling and expansion rate}
The simplicity of the symmetrized figures (figs.~\ref{f5}-\ref{f6}) is in contrast with the more complex real 3D structure - and the physics that is really at work.
It not only gives a false impression of the real symmetries, but it also suggests that the anisotropy is scale-independent, which, strictly, cannot be true - and is not true.
Indeed, the expansion timescale is independent of scale, while the nonlinear
timescale and the relative fluctuations' amplitude, $b_k/B_0$, decrease with scale. 
Since the latter two control the amount of anisotropy with respect to the mean
field, we also expect the anisotropy to depend on scale.
Indeed, if we look at fig.~\ref{f3}f (run C, isotropic initial conditions), one sees that the actual symmetry axis changes systematically when going from large scales to small scales.

Thus, varying the expansion parameter should allow us to vary the extent of the
scales dominated by expansion and those dominated by nonlinear couplings. 
We now consider the two series or runs, runs $\mathrm{D_{1...5}}$ and
$\mathrm{E_{1...5}}$ with increasing expansion parameter
$\epsilon$ and different inital condition (\gyrotropic{} and isotropic, 
respectively).
To describe the change of anisotropy with scale in a given run, we measure for each
isocontour of the 2D ecliptic spectrum the maximal distance from the origin
$k_L=\sqrt{k_x^2+k_y^2}$ and the corresponding angle with respect to the radial
direction $\theta_R$ (an illustration of the method is given in
fig.~\ref{f3}f).
The result is shown in fig.~\ref{f7}a for runs with \gyrotropic{} initial
conditions, and in fig.~\ref{f7}b for runs with isotropic initial conditions
(see Table~\ref{table1}).
\chb{Note that for run A, $k_L$ is renormalized by a factor 5, in order to allow compare run A without expansion with runs with expansion which have different domain sizes (see fig.~\ref{f3}).}

Consider first the series, $\mathrm{D_{1...5}}$, with \gyrotropic{} initial conditions
and decreasing $\epsilon$, fig.~\ref{f7}a.
In absence of expansion the standard perpendicular cascade (that is, perpendicular to $B_0$) 
is well measured by the method, with $\theta_R\sim-45^o$ at small enough scales (solid line).
With expansion, the principal-axis angle $\theta_R$ slightly decreases with $k$. For large $k$ 
and large expansion parameter it is clustered around $-37^o$ 
and it approaches $-42^o$ at the smallest expansion parameter $\epsilon=0.2$, that
is the $B_0$-anisotropy.
This suggests that the \gyrotropy{} is an attractor at small scales
for turbulence in the expanding solar wind (recall that smaller expansion
parameters correspond roughly to smaller scales for given solar wind speed
and fluctuations' amplitude, see the definition of $\epsilon$ in
eq.~\eqref{eps}).


Consider now the series of runs $\mathrm{E_{1...5}}$ with isotropic initial
conditions, fig.~\ref{f7}b. 
With expansion, all angles $\theta_R$ systematically decrease with $k$, with
possibly a common asymptote at $-34^o$ as indicated by runs with
$\epsilon\in[0.1,~0.4]$.
Whether the true asymptote is the $B_0$-anisotropy (i.e., $\theta_R=-45^0$) 
or an intermediate value, as suggested by the figure, is to be proven with simulations made at much larger Reynolds numbers. 

Whatever the true asymptotic value, the important result is that, with
reasonable values of $\epsilon_{end}$ at 1~AU, that is between $0.3$ and $2$
\citep{1991AnGeo...9..416G}, the first decade of the inertial range has an
anisotropy that is strongly influenced by expansion since the 
symmetry axis is determined by \textit{both} the radial and the mean-field direction.

\begin{figure}[t]
\begin{center}
\includegraphics [width=0.48\linewidth]{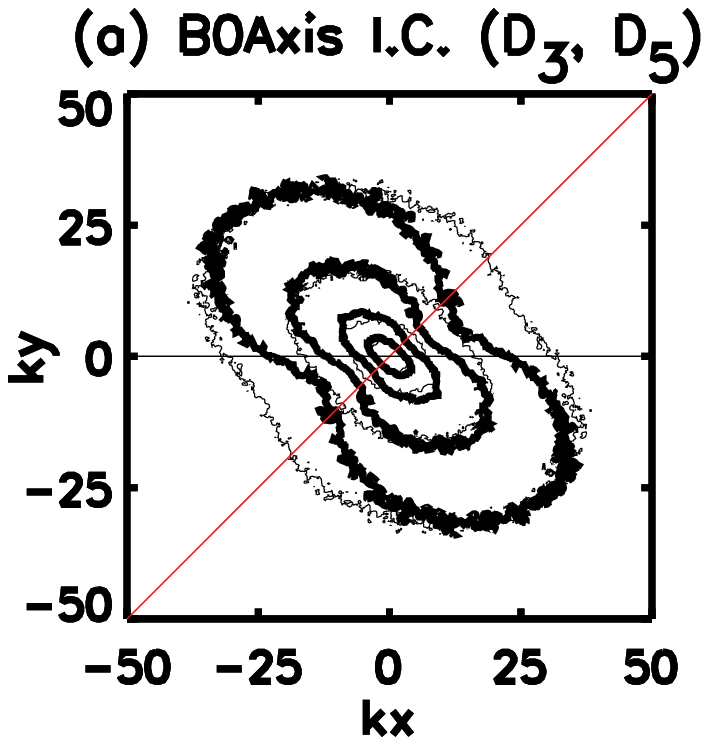}
\includegraphics [width=0.48\linewidth]{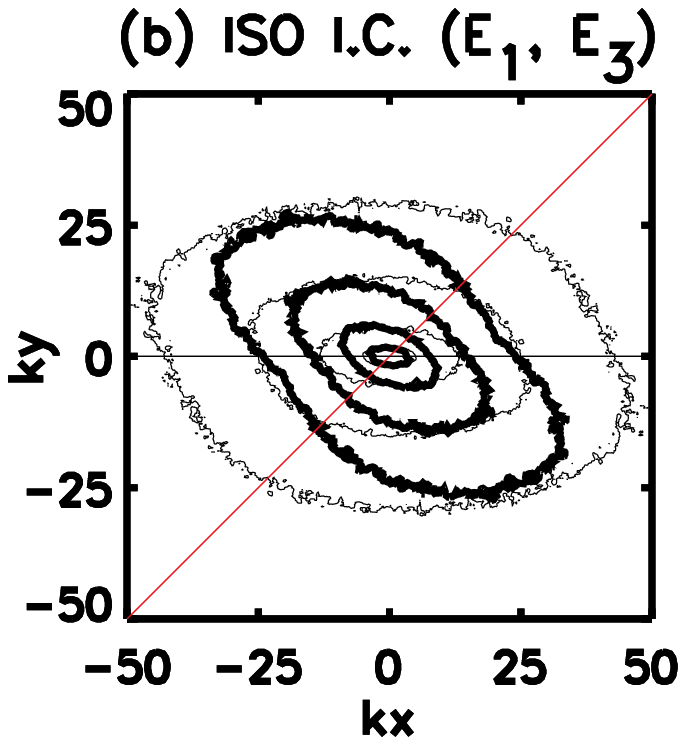}
\caption{Superposition of final 2D isocontours (ecliptic cuts, $k_z=0$) for two values of
$\epsilon$: 0.4 (thick lines) and 2 (thin lines).
(a) \gyrotropic{} initial conditions; (b) isotropic initial conditions.
}
\label{f8}
\end{center}
\end{figure}

The insensitivity to expansion of the runs with \gyrotropic{} initial conditions compared
to isotropic initial conditions is illustrated in fig.~\ref{f8}. We overplot in
each case the isocontours of two runs with $\epsilon=0.2$ (thick lines) and
$\epsilon=2$ (thin lines). For \gyrotropic{} initial conditions the final contours are almost superposed, while with isotropic initial conditions the final contours 
vary strongly with $\epsilon$:
the isocontours with large $\epsilon$ are almost isotropic (actually compatible
with $\theta_R \simeq 0)$, while with small $\epsilon$ they are rather close to
the isocontours with initial \gyrotropy.

When expansion matters, as in the case of isotropic spectra,
the final anisotropy keeps trace of the particular initial spectrum. Consider run C in
fig.~\ref{f3}f and run $\mathrm{E_1}$ in fig.~\ref{f8}b (thick lines), which differ only
by the value of the cut in the initial spectrum, $k_{cut}^y=128,~64$, respectively.
Expansion causes the same slow down of nonlinear interactions but the final
spectrum is qualitatively different, with run C showing a stronger symmetry
around the radial axis. This is because in run C we initially excited modes
$k_y>k_x$ while in run $\mathrm{E_1}$ the spectrum is truly isotropic.
All modes at high $k_y$ undergo a kinematic contraction in the Fourier space
independently of their $k_x$ extent,
so the differences in the final anisotropy arise from the freezing-in of the
different intial spectrum.


\section{Conclusions}
We studied the anisotropy of turbulence in the solar wind carrying out
numerical simulations of the expanding box model for MHD (EBM). 
We varied both the initial conditions and the expansion rate of our simulations,
thus extending recent works on the evolution of turbulence in the solar wind
\citep{Dong:2014fi}. To compare with solar wind data we computed how the anisotropy 
shows up in 2D autocorrelation functions.

We found that if the initial spectrum
is already axisymmetric with respect to the mean field, then the spectrum at 1~AU 
conserves this symmetry and shows up as a 2D-turbulence component, consistent
with the above assumption.
However, if the initial spectrum is isotropic, then 
the spectrum at 1~AU is not axisymmetric and the anisotropy is determined
by two symmetry axes, the radial axis and the mean-field axis. 
This is true for a large range of expansion rates and of wavenumbers, suggesting that
the mean-field anisotropy is a weak attractor, or in other words that the
recovery of homogenous-turbulence properties at small scales is not a universal feature
of solar wind turbulence. 

We also showed that the assumption of axisymmetry about the mean field, which is
often conjectured to hold at small enough scales, may mask the true anisotropy of
the magnetic field spectrum. 
In fact, when the spectrum displays a slab component along the radial, 
as for isotropic initial conditions, the assumption of axisymmetry about the mean field  
transforms the anisotropy into an apparent slab component along the mean field. 

Thus, on the one hand we confirm earlier results of homogenous turbulence simulations 
for the origin of the 2D component \citep{1990JGR....9520673M,1998JGR...10323705G}, on the other hand we provide an explanation for the slab component observed in
fast streams. 

What controls the anisotropy at 1~AU?
The \gyrotropic{} initial conditions we have used do not only possess the ``right'' symmetry
properties, but also an aspect ratio that is characteristic of strong
turbulence (compare the values of $\chi$ in table~\ref{table1}). For reasonable expansion rates, $\epsilon_{end}\in[0.4,~2]$ \citep{1991AnGeo...9..416G}, turbulence remains strong and its properties are similar to homogenous
turbulence. 
On the contrary, in isotropic initial conditions we excited a large range of field-parallel
wavevectors, which makes the cascade weaker. On top of this, expansion slows
down the nonlinear interaction due to the kinematic stretching of the plasma. 
Thus, we have two weakening factors that counteract the natural tendency of MHD
turbulence to develop small scales perpendicular to the mean field. 
In this case, the development of turbulence is more sensitive to the expansion rate
and the final anisotropy depends on the expansion symmetry axis, the
radial, and on the initial anisotropy.

This leads us to conjecture that slow-wind turbulence is already strongly anisotropic
with symmetry axis given by the mean field, and that fast-wind turbulence is
more isotropic.

Recall that in fast-wind turbulence a strong correlation between velocity and
magnetic fluctuations is observed (high cross-helicity), which results into an additional weakening of the cascade 
\citep{2012ApJ...750L..33V,Perez:2013fb}. Such weakening could also be
responsible for the formation of the 1/f spectrum 
inside the Alfv\'enic critical point \citep{2012ApJ...750L..33V}, which is a
characteristic of the fast solar wind \citep[e.g.][]{Bruno:2013p434}.
Preliminary EBM simulations with initial strong 
cross helicity and isotropic spectra compare well
with turbulence observed in the fast streams \citep{1990JGR....95.8197G}. Indeed,
at 1~AU they have flatter spectra and a higher cross helicity compared to runs
with \gyrotropic{} initial
spectra, suggesting that one needs to account for all the three
factors (expansion rate, initial anisotropy, and initial cross helicity) in order to
understand the different evolution of turbulence in fast and slow streams.

\chb{The so-called Bieber test \citep{1996JGR...101.2511B,Saur:1999gy,Smith:2011ep} that relates spectral and component anisotropy 
could be used to further test our conjecture after a proper generalization, i.e., by including the radially-symmetric models of turbulence proposed here.}

We conclude by noting that the NASA Solar Probe Plus and ESA Solar Orbiter missions will
sample plasma in between 0.1~AU and 0.8~AU. This makes extremely interesting and
timely to understand which mechanisms can lead to different initial anisotropies close to the Sun
for fast and slow streams.
Shell-Reduced-MHD simulations 
\citep{2009ApJ...700L..39V,2012A&A...538A..70V,2012ApJ...750L..33V,2012PhRvL.109b5004V}
are particularly promising, since allow to span 5 decades in wavenumbers and 
the large parameter space that characterizes slow and fast wind, as well as true Reduced MHD simulations
\citep{Perez:2013fb}, since they provide more detailed informations on
turbulence. Another promising tool is the Accelerating Box Model \citep{2013JGRA..118.7507T}, 
which not only incorporates the acceleration of the solar wind into the EBM
but also allows to treat compressible effects, such as parametric
instability \citep[e.g][]{2015JPlPh..81a3202D}, that are neglected in the
above models and may contribute to the acceleration of the solar wind and shape the turbulent spectrum
close to the Sun \citep{Suzuki:2005kf,Matsumoto:2012ck}.\\

\textit{Acknowledgments} 
We acknowledge R. Bruno, W. H. Matthaeus, and T. K. Suzuki for useful discussions on an earlier version of the manuscript. This work has been done within the LABEX PLAS@PAR project, and received
financial state aid managed by the Agence Nationale de la Recherche, as part of
the Programme ``Investissements d'Avenir'' under the reference
ANR-11-IDEX-0004-02.
HPC resources were provided by CINECA (grant 2015 HP10CVTCYK)
\chb{and by GENCI-IDRIS (grant 2016 047683)}.

\bibliographystyle{apj}

\end{document}